\documentclass[osajnl,twocolumn,showpacs,superscriptaddress,10pt]{revtex4-1} 
\usepackage{amsmath,amssymb,graphicx}
\begin{document}

\title{Coupled-mode induced transparency in aerostatically-tuned microbubble whispering gallery resonators}

\author{Yong Yang}\email{Corresponding author:yong.yang@oist.jp}
\author{Sunny Saurabh}
\affiliation{Light-Matter Interactions Unit, Okinawa Institute of Science and Technology, Onna, Okinawa 904-0495, Japan}
\author{Jonathan Ward}
\author{S\'ile Nic Chormaic}
\affiliation{Light-Matter Interactions Unit, Okinawa Institute of Science and Technology, Onna, Okinawa 904-0495, Japan}

\begin{abstract}Coupled-mode induced transparency is realized in a single microbubble whispering gallery mode resonator. Using aerostatic tuning, we find that the pressure induced shifting rates are different for different radial order modes. A finite element simulation considering both the strain and stress effects shows a GHz/bar difference and this is confirmed by experiments. A transparency spectrum is obtained when a first order mode shifts across a higher order mode through precise pressure tuning.  The resulting lineshapes are fitted with the theory. This work lays a foundation for future applications in microbubble sensing.
\end{abstract}

\maketitle 
In the last few decades, a category of optical microcavities, termed whispering gallery mode (WGM) resonators (WGRs), has been developed and has generated interest in a wide range of fields.  These resonators are characterized by having  ultrahigh quality (Q) factors and small optical mode volumes, thereby making them advantageous for studies in areas such as cavity QED \cite{Aoki2006,Park2006}, nonlinear optics \cite{Herr2013}, optomechanics \cite{Aspelmeyer2013} and high sensitivity sensing \cite{Vollmer2008}. Within the WGR family, the microbubble resonator \cite{Sumetsky2010} is a relatively recently-developed type. Experimentally, microbubbles can be made with single \cite{Watkins2011} or double inputs \cite{Sumetsky2010} so that different materials can be injected into the hollow core. A subtle change in the material can cause a large frequency shift to the resonator's modes, which can be used for sensing. Extensive work has already be done in the field of bio \cite{Li2013,Sun2011} and chemical sensing \cite{Sun2011} using WGMs in hollow optical resonators. Various microbubble sensing applications have been studied, such as refractive index \cite{White2006}, temperature \cite{Ward2013}, and optomechanical sensing \cite{Bahl2013}.  Aside from sensing, lasing in hollow cavities has been achieved \cite{Suter2010,Lee2011}.

Another potential application of the microbubble is that of coupled-mode induced transparency. Electromagnetic induced transparency (EIT)-like and Fano-like lineshapes in the transmission spectrum of a WGM resonator coupled to a waveguide, such as a tapered optical fiber, have been observed in recent years. These phenomena are very important for applications such as slow light \cite{Totsuka2007} and optical switching \cite{Fan2002}. To date,  EIT- and Fano-like lineshapes have been seen in a single microtoroid \cite{Xiao2009,Li2011}, a single microsphere \cite{Dong2009}, and in directly \cite{Zheng2012} or indirectly coupled \cite{Li2012} resonators. For a single resonator, a low and a high Q mode should be found  within a very small  frequency separation. When the two modes are tuned into resonance, an EIT-like lineshape occurs.

In this paper we will show that resonance can be achieved by pressure tuning a single microbubble resonator \cite{Henze2011}, so that two modes with different mode orders experience different shift rates. Aerostatic tuning of the modes has some advantages over other methods. For example, it is precise and fast, with a larger tuning range than dispersion tuning in a single microsphere \cite{Dong2009}. Also, it is more convenient and repeatable than temperature tuning \cite{Ward2013}.

WGMs with different orders have different optical paths, or different effective mode radii, in the cavity. Therefore, modes of different orders show different FSRs and responses to external perturbations, such as aerostatic pressure. These conditions are necessary for the EIT phenomena reported in this paper, because they lead to modes with close frequencies and varying aerostatic tuning rates. Consider a microbubble of inner and outer radius, $b$ and $a$, respectively. A microbubble with a thick wall, i.e., $(a-b)>>\lambda$, can be treated equivalent to a microsphere, at least for the radial modes $q=1$ and 2 when the resonances are determined as an asymptotic formula \cite{Lam1992}. The sensitivity is given as
\begin{equation}
\begin{split}
\frac{d\lambda}{\lambda}= &\frac{da}{a}+\left\{\right.1-\frac{\lambda}{2\pi a}[\frac{1}{(n^2-1)^{1/2}}-\frac{n^2}{(n^2-1)^{3/2}}-\\
& Ai_q\nu^{-2/3}[\frac{n^2}{(n^2-1)^{3/2}}-\frac{n^4}{(n^2-1)^{5/2}}]]\left. \right\}\frac{dn}{n},
\end{split}
\label{sens}
\end{equation}
where $n$ is the refractive index of the microbubble, $\nu=l+1/2$ and $l$ is the angular momentum of the resonance. $Ai_q$ is the $q^{th}$ root of the Airy function, $Ai(-z)$.
Note that in \cite{Henze2011} high order terms in the asymptotic expression were omitted so that $d\lambda/\lambda\approx da/a+dn/n$.
For the two modes, $q=1$ and 2, with the same $\lambda_0$, the shift rates differ by
\begin{equation}
\begin{split}
\triangle(\frac{d\lambda}{\lambda})& =\frac{\lambda_0}{2\pi a}\frac{dn}{n}[\frac{n^2}{(n^2-1)^{3/2}}-\frac{n^4}{(n^2-1)^{5/2}}]\times\\
& [Ai_1\nu_1^{-2/3}-Ai_2\nu_2^{-2/3}].
\label{sensdiff}
\end{split}
\end{equation}

Here, $Ai_1<Ai_2$ and, for a fixed $\lambda_0$, $l_1>l_2$ in general, so that $Ai_1\nu_1^{-2/3}<Ai_2\nu_2^{-2/3}$. For a silica microbubble, $n=1.4457$. Eq. \ref{sensdiff} is always positive, therefore low order modes shift faster than higher ones. The larger the value of $dn/n$, the larger the shift rates between the modes. From \cite{Henze2011}, $dn/n$ is proportional to a geometrical parameter, $\chi=b^3/(a^3-b^3)$, and we can write
\begin{equation}
\frac{dn}{n}=\frac{3(p_ib^3-p_oa^3)}{n(a^3-b^3)}C\approx\frac{3\chi C(p_i-p_o)}{n},
\label{stress}
\end{equation}
where $C=4\times 10^{-12}$ m$^2$/N is the elasto-optic constant of silica, and $p_i$ and $p_o$ are the internal and external pressures. Therefore, for a thinner wall, the effect should be more obvious. However, the earlier assumption that the mode distribution is equivalent to that of a microsphere is not valid for thin-walled microbubbles; in the following discussion, we adapt a finite element method \cite{Yang2014} to precisely simulate the microbubble modes. The fact that the strain contribution will lead to a much larger shift rate difference in thin-walled bubbles should also be considered and we will discuss this later.

\begin{figure}[htbp]
\centerline{\includegraphics[width=\columnwidth]{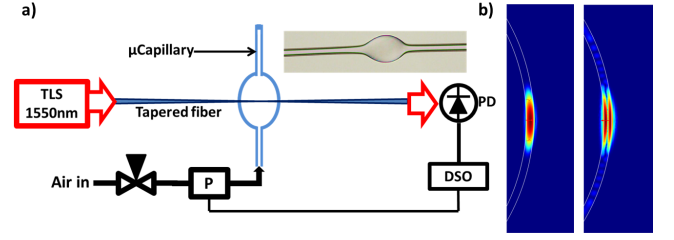}}
\caption{a) Schematic of the experiment. TLS: tunable laser source; DSO: digital oscilloscope; PD: photon detector. P is an electrical pressure sensor. The inset is the photo of the microbubble used in the experiments, with a diameter of 90 $\mu$m and wall thickness of 1.26 $\mu$m. b) The simulated EM field distribution for the $q=1$ and 2 modes (from left to right) using a finite element method.}
\label{scheme}
\end{figure}

\begin{figure}[htbp]
\centerline{\includegraphics[width=\columnwidth]{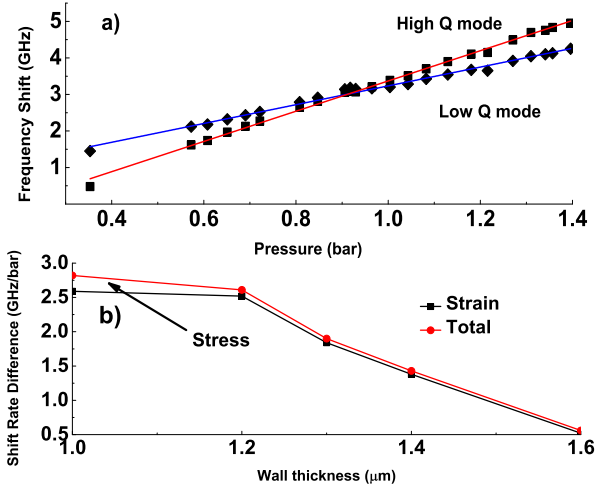}}
\caption{(a) Experimentally measured pressure tuning rate of a high Q mode (squares) and a low Q mode (spades) in a microbubble. The solid red and blue lines are linear fits to the data. (b)Frequency shift rate differences of $q=1,2$ modes for 1-1.6 $\mu$m thicknesses. The black squares are due to the strain effect and red circles are total rate differences. Lines are only for guiding the eye.}
\label{tune}
\end{figure}

A microbubble was made using the method described in \cite{Ward2013} with one input sealed using a CO$_2$ laser. The radius and wall thickness were measured as $a=45$ $\mu$m and $t=a-b=1.26$ $\mu$m. The WGM shifts were recorded using a taper-coupled  setup illustrated in Fig. \ref{scheme}(a) and an electronic sensor was calibrated for monitoring the internal pressure. The taper was made by heating-and-pulling standard 1550 nm single-mode fiber down to a waist  of less than 1.2 $\mu$m to ensure single mode operation. To eliminate the influence of vibrations, the microbubble was in contact coupling with the tapered fiber and the whole system was placed in an enclosure. Two modes with different Q factors were chosen for tracking. It is assumed that the low Q mode was $q=2$ and the high Q mode was $q=1$. Resonant frequencies at different internal pressures are plotted in Fig. \ref{tune}(a). The linear fits yield  shift rates of $4.27$ GHz/bar and $2.60$ GHz/bar for the two modes, resulting in a positive rate difference of $1.67$ GHz/bar.

To simulate the experimental results, the frequency shifts due to stress and strain were calculated separately.  By setting $p_i-p_o$ in Eq. \ref{stress} to 0, $1\times10^4$ Pa (i.e., 10 bar) and $2\times10^4$ Pa (i.e., 20 bar) we numerically calculated the frequency shifts for the $q=1$ and 2 modes (Fig \ref{scheme}(b)) in a bubble with a diameter of 90 $\mu$m and a wall thickness varying  from 1 $\mu$m to 1.6 $\mu$m. The stress effect, $dn/dp$, causes a shift rate difference in the MHz/bar scale only.

Strain affects the WGM resonance by changing the size of the microbubble as $da/dp$. FEM simulations were used to calculate the shifts for the $q=1$ and $2$ modes for a given size expansion (i.e., 0.01 $\mu$m increase in $a$). Next, the relative change in $a$, given by \cite{Henze2011}
\begin{equation}
\frac{da}{a}\approx\frac{(4G+3K)\chi}{12GK}(p_i-p_o),
\label{strain}
\end{equation}
was used to determine the required pressure for the same expansion. This was repeated for different wall thicknesses. The shear and bulk moduli in Eq.\ref{strain} are $G=31\times10^9$ Pa and $K=41\times10^9$ Pa for the silica capillary used.
As a summary, Fig \ref{tune}(b) illustrates the shift rate differences between the two modes originating from the strain and the stress, thus giving the total rate difference. For thin-walled microbubbles the strain-induced rate difference is in the GHz/bar range. The shift rate difference for the $q=1$ and 2 modes for a wall thickness of 1.3 $\mu$m is calculated to be 1.90 GHz/bar, which is quite close to the  measured value from Fig. \ref{tune}(a).

\begin{figure}[htbp]
\centerline{\includegraphics[width=\columnwidth]{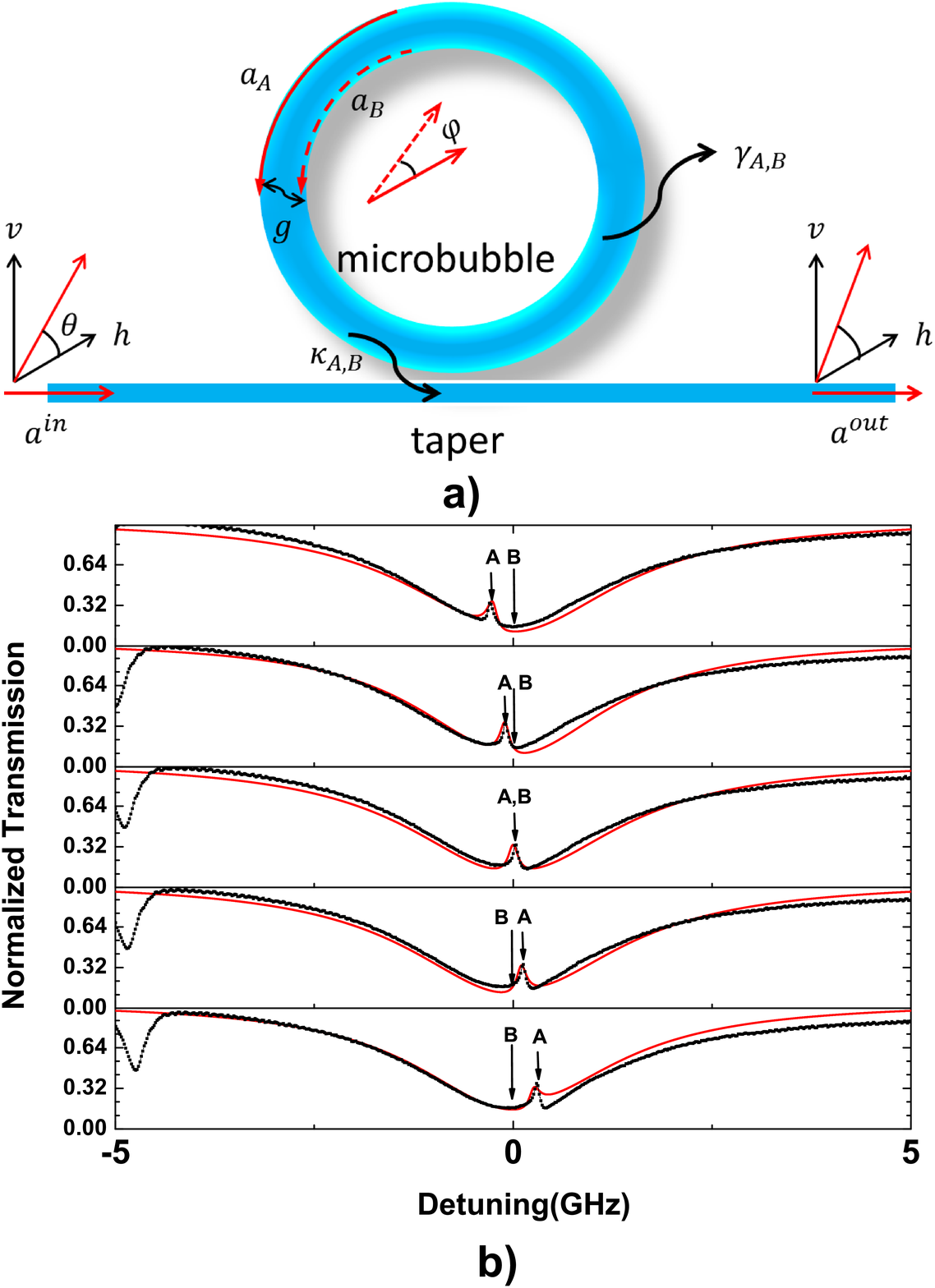}}
\caption{(a) A schematic illustration of coupled mode model of a microbubble coupled with taper. (b) Fitted EIT-like shape with pressure tuning. The red curves are theoretical fits with fitting parameters [$\kappa_A,\gamma_A,\kappa_B,\gamma_B] =[0.1,0.08,1.0,2.2]$ GHz,  $\varphi=\pi/10$ and $\theta=\pi/6$. From top to bottom the detunings are 0.25, 0.1, 0, -0.1 and -0.25 GHz.}
\label{eit}
\end{figure}

In experiments, the linewidth of the low Q mode is about 2 GHz. We could always find  a high Q mode (linewidth of MHz) in the vicinity of the low Q one. With only about 1 bar of input pressure, the high Q mode (A) can be tuned across the low Q mode (B) transmission dip. Fig. \ref{eit}(b) shows an example of such a crossing, where an EIT-like shape appears as a result of tuning the two modes into resonance.

In \cite{Li2011}, the appearance of EIT and Fano shapes in a single microresonator was explained by coupled-mode theory as follows:
\begin{equation}
\begin{split}
\frac{da_A}{dt} &=-i\omega_Aa_A-\frac{\gamma_A+\kappa_A}{2}a_A-ga_B+\sqrt{\kappa_A}a^{in}\cos{\theta}.\\
\frac{da_B}{dt} &=-i\omega_Ba_B-\frac{\gamma_B+\kappa_B}{2}a_B-ga_A+\\
&\{\sqrt{\kappa_B}\cos{\varphi}a^{in}\cos{\theta}
+\sqrt{\kappa_B}\sin{\varphi}a^{in}\sin{\theta}\}.
\end{split}
\label{CMT}
\end{equation}
A and B are considered to be two quasi TE (TM) WGMs indirectly coupled through the taper mode $a^{in}$. $\gamma_{A,B}$ are the intrinsic rates and $\theta$ is the polarization orientation of the taper mode $a^{in}$, see Fig. \ref{eit}(a). The coupling strength is determined by the external coupling rates, $\kappa_A$ and $\kappa_B$, and their polarization angles, $\varphi$, through $g=(\sqrt{\kappa_A\kappa_B}\cos{\varphi})/2$. Intracavity EM fields of the A, B modes, i.e. $a_A$ and $a_B$, are coupled back to the taper fiber mode, with a different input/output relationship for two orthogonal polarizations as $a^{out}_{h}=a^{in}\cos{\theta}+\sqrt{\kappa_A}a_A+\sqrt{\kappa_B}\cos{\varphi}a_B$ and $a^{out}_v=a^{in}\sin{\theta}+\sqrt{\kappa_B}\sin{\varphi}a_B$. The transmission spectrum is an intensity superposition of these to orthogonal polarizations:
\begin{equation}
T=\left|\frac{a^{out}}{a^{in}}\right|^2=\frac{|a^{out}_h|^2+|a^{out}_v|^2}{|a^{in}|^2}.
\label{Trans}
\end{equation}
From Eq. \ref{CMT}, $a_{A,B}$ can be deduced when in steady state, i.e. $da/dt=0$. By substituting into Eq. \ref{Trans}, the transmission can be calculated for known $[\kappa_A,\kappa_B,\gamma_A,\gamma_B,\varphi,\theta]$ and detuning $\triangle\omega=\omega_B-\omega_A$. Thus EIT and Fano-like lineshapes can be manipulated by either changing the detuning or polarization \cite{Li2011,Dong2009}.

In contrast to  earlier work \cite{Dong2009}, where the detuning of the two modes is modified by varying the coupling gap and, subsequently, by dispersion, our system remains in contact with the coupling fiber. The pressure tuning only changes the detuning of the modes with a much larger range, while keeping other parameters unchanged. Since the rate difference has been determined in the previous discussion, the detuning can be precisely known if the pressure is also known. In practice, the experimental data was fitted with the detuning ranging from 250 MHz to -250 MHz and using one unchanged set of parameters, $[\kappa_{A,B},\gamma_{A,B},\varphi,\theta]$, which are given in Fig. \ref{eit}. Note there is still a small discrepancy when adding more pressure (i.e. detuning is more negative). A possible explanation is that silica is a birefringent material, so pressurizing the microbubble may rotate the polarization of the WGMs, thereby distorting the lineshape from EIT-like to Fano-like.

In summary, we have shown that a controllable EIT-like shape can be easily realized in a single aerostatically-tuned microbubble. We proved, both theoretically and experimentally, that, for different radial modes, the pressure tuning rates are different due to the size and refractive index change induced by strain and stress. The experiments show that aerostatic tuning of the microbubble  might be a simplified and precise way to implement controllable EIT and Fano lineshapes. Furthermore, by having a Fano-like lineshape due to the coupling of two modes, the sensitivity for pressure sensing in a microbubble can be enhanced, as predicted in \cite{Li2011}.

This work was supported by the Okinawa Institute of Science and Technology Graduate University. The authors thank N. Dhasmana for initial contributions to this work.

\end{document}